%% file: main.tex
\documentclass[sigconf,screen]{acmart}
\AtBeginDocument{%
  \providecommand\BibTeX{{%
    \normalfont B\kern-0.5em{\scshape i\kern-0.25em b}\kern-0.8em\TeX}}}

\setcopyright{acmcopyright}
\copyrightyear{2022}
\acmYear{2022}
\acmDOI{10.1145/1122445.1122456}

\acmConference[ISSTA 2022]{ACM SIGSOFT International Symposium on Software Testing and Analysis}{18-22 July, 2022}{Daejeon, South Korea}
\acmBooktitle{ISSTA '22: ACM SIGSOFT International Symposium on Software Testing and Analysis, 18-22 July, 2022, Daejeon, South Korea}
\acmPrice{15.00}
\acmISBN{978-1-4503-XXXX-X/18/06}

\usepackage{multirow}
\usepackage{multicol}
\usepackage{enumitem}

\usepackage{xcolor}
\usepackage{colortbl}
\usepackage{tcolorbox}
\definecolor{mygray}{RGB}{226,226,226}

\usepackage{algorithm}
\usepackage{algpseudocode}

\usepackage{balance}

\newcommand{\methodname}{AEON}

\begin{document}

\title{{\methodname}: A Method for Automatic Evaluation of NLP Test Cases}

\author{Jen-tse Huang}
\affiliation{
  \institution{The Chinese University of Hong Kong}
  \city{Hong Kong}
  \country{China}
}
\email{jthuang@cse.cuhk.edu.hk}

\author{Jianping Zhang}
\affiliation{
  \institution{The Chinese University of Hong Kong}
  \city{Hong Kong}
  \country{China}
}
\email{jpzhang@cse.cuhk.edu.hk}

\author{Wenxuan Wang}
\affiliation{
  \institution{The Chinese University of Hong Kong}
  \city{Hong Kong}
  \country{China}
}
\email{wxwang@cse.cuhk.edu.hk}

\author{Pinjia He}
\authornote{Corresponding author.}
\affiliation{
  \institution{The Chinese University of Hong Kong, Shenzhen}
  \city{Shenzhen}
  \country{China}
}
\email{hepinjia@cuhk.edu.cn}

\author{Yuxin Su}
\affiliation{
  \institution{Sun Yat-sen University}
  \country{China}
}
\email{suyx35@mail.sysu.edu.cn}

\author{Michael R. Lyu}
\affiliation{
  \institution{The Chinese University of Hong Kong}
  \city{Hong Kong}
  \country{China}
}
\email{lyu@cse.cuhk.edu.hk}

\renewcommand{\shortauthors}{Huang and Zhang, et al.}

\begin{abstract}
Due to the labor-intensive nature of manual test oracle construction, various automated testing techniques have been proposed to enhance the reliability of Natural Language Processing (NLP) software.
In theory, these techniques mutate an existing test case (\textit{e.g.}, a sentence with its label) and assume the generated one preserves an equivalent or similar semantic meaning and thus, the same label.
However, in practice, many of the generated test cases fail to preserve similar semantic meaning and are unnatural (\textit{e.g.}, grammar errors), which leads to a high false alarm rate and unnatural test cases.
Our evaluation study finds that 44\% of the test cases generated by the state-of-the-art (SOTA) approaches are false alarms.
These test cases require extensive manual checking effort, and instead of improving NLP software, they can even degrade NLP software when utilized in model training.
To address this problem, we propose {\methodname} for \textbf{A}utomatic \textbf{E}valuation \textbf{O}f \textbf{N}LP test cases.
For each generated test case, it outputs scores based on semantic similarity and language naturalness.
We employ {\methodname} to evaluate test cases generated by four popular testing techniques on five datasets across three typical NLP tasks.
The results show that {\methodname} aligns the best with human judgment.
In particular, {\methodname} achieves the best average precision in detecting semantic inconsistent test cases, outperforming the best baseline metric by 10\%.
In addition, {\methodname} also has the highest average precision of finding unnatural test cases, surpassing the baselines by more than 15\%.
Moreover, model training with test cases prioritized by {\methodname} leads to models that are more accurate and robust, demonstrating {\methodname}'s potential in improving NLP software.
\end{abstract}



\maketitle

\input Sections/1_Introduction.tex
\input Sections/2_Preliminaries.tex
\input Sections/3_Approaches.tex
\input Sections/4_Settings.tex
\input Sections/5_Results.tex
\input Sections/6_Related_Work.tex
\input Sections/7_Conclusion.tex

\begin{acks}
We thank gratefully the constructive discussion with Shuqing Li, and the helpful feedback from members of ARISE Lab.
The work described in this paper was supported by the key program of fundamental research from the Shenzhen Science and Technology Innovation Commission (No. JCYJ20200109113403826), the Research Grants Council of the Hong Kong Special Administrative Region, China (CUHK 14210920 of the General Research Fund), and the National Natural Science Foundation of China (No. 62102340).
\end{acks}

\balance

\bibliographystyle{ACM-Reference-Format}
\bibliography{reference}

\end{document}

%% file: Sections/1_Introduction.tex
\section{Introduction}
\label{sec-introduction}

NLP software has become increasingly popular in our daily lives.
For example, NLP virtual assistant software, such as Siri and Alexa, receives billions of requests \cite{siri2018, Alexa2021} while Google Translate App translates more than 100 billion words per day \cite{translation2016}.
With the development of Deep Neural Networks (DNNs), the performance of NLP software has been largely boosted.
Equipped with the SOTA model \cite{transformer}, Microsoft question answering robot surpasses humans on conversational question answering task.
In addition, the performance of machine comprehension \cite{yifan1}, text generation \cite{jingjing1} and machine translation \cite{wenxuan1} has been significantly improved.
However, NLP software can produce erroneous results, leading to misunderstanding, financial loss, threats to personal safety, and political conflicts \cite{problem1, problem2}.

\begin{table*}
\caption{
  Examples for high-quality, inconsistent, and unnatural test cases generated by existing testing techniques on different datasets.
  NLP tasks include Sentiment Analysis (SA), Natural Language Inference (NLI), and Semantic Equivalence (SE).
  Mutated words are marked in \textcolor{red}{red}.
}
\label{tab-cover}
\begin{center}
\begin{tabular}{l|cc}
\toprule
\bf Original text & \bf Task & \bf Technique \\
\bf Generated test case & \bf Dataset & \bf Issue \\
\cmidrule(lr){1-3}
A man under a running shower with shampoo in his hair. $\Rightarrow$ A man is taking a \textcolor{red}{shower}. & NLI & BAE \cite{bae} \\
A man under a running shower with shampoo in his hair. $\Rightarrow$ A man is taking a \textcolor{red}{bath}. & SNLI \cite{snli} & \bf None \\
\cmidrule(lr){1-3}
Ultimately this is a \textcolor{red}{frustrating} patchwork. & SA & PSO \cite{pso} \\
Ultimately this is a \textcolor{red}{sparkling} patchwork. & MR \cite{mr} & \bf Inconsistent \\
\cmidrule(lr){1-3}
\textcolor{red}{British} action wouldn't have mattered. $\Rightarrow$ British action would have made a big difference. & NLI & BAE \cite{bae} \\
\textcolor{red}{Welsh} action wouldn't have mattered. $\Rightarrow$ British action would have made a big difference. & MNLI \cite{mnli} & \bf Inconsistent \\
\cmidrule(lr){1-3}
What are some good topics to \textcolor{red}{be} bookmarked on Quora? & SE & Textfooler \cite{textfooler} \\
What are some good topics to \textcolor{red}{es} bookmarked on Quora? & QQP \cite{glue} & \bf Unnatural \\
\cmidrule(lr){1-3}
\textcolor{red}{I went} to \textcolor{red}{Danny's} this weekend to \textcolor{red}{get} an oil change and car wash and \textcolor{red}{I paid} for a VIP car wash. & SA & BAE \cite{bae} \\
\textcolor{red}{My gone} to \textcolor{red}{work} this weekend to \textcolor{red}{do} an oil change and car wash and \textcolor{red}{my hired} for a VIP car wash. & Yelp \cite{agnews-yelp} & \bf Unnatural \\
\bottomrule
\end{tabular}
\end{center}
\end{table*}

To discover erroneous behaviors in NLP software, researchers have designed various software testing techniques \cite{pinjia1, checklist, textbugger, pso, chen2021validation}.
A test case for NLP software is in the form of a text (\textit{e.g.}, a sentence) and its label, where the label is the expected correct output of the NLP software.
In theory, most of these testing techniques modify part(s) of the input text (\textit{e.g.}, a word/character substitution/insertion/deletion) under the assumption that the generated test case preserves an equivalent or similar semantic meaning.
Typically, these techniques take labeled texts as inputs and output the mutated texts and the corresponding labels.

However, it is still challenging for current testing techniques to produce practical test cases of high quality.
Specifically, tiny modification in a text can change its semantic meaning, which invalidates the common assumption that the semantic meaning of the original text and that of the generated text should remain equivalent or similar, further rendering the possibility of changing the corresponding labels \cite{reevaluate1, reevaluate2}.
For example, removing ``not'' from the text ``I do not like the movie'' changes its semantic meaning and further changes its label for a sentiment analysis task from ``negative'' to ``positive'', resulting in a test case with an incorrect label and further a false alarm.
Moreover, existing testing approaches cannot guarantee the fluency and naturalness of the generated test cases.
Many word-level testing approaches introduce grammar errors and punctuation errors, and sometimes they introduce words that do not exist or are rarely used~\cite{reevaluate1}.
Although these test cases may trigger ``software errors'' (\textit{e.g.}, unexpected software behaviors),  it is important to first ensure the quality of the test cases in terms of semantic consistency and naturalness before finding more errors.

According to our user study, many of the NLP test cases generated by existing approaches are of low quality because of the following two issues: \textit{Inconsistent} issue and \textit{Unnatural} issue.
These issues can lead to false alarms in testing and unnaturalness in language.
In this paper, we say an NLP test case is of high quality if it does not have any of these issues.
As shown in Table~\ref{tab-cover}, a high-quality test case preserves the semantics of the original text and reads smoothly.
The first \textit{Inconsistent} case changes the semantics to the opposite while the second one changes the subjects.
Two \textit{Unnatural} cases hurt the fluency and naturalness of natural language by introducing either non-existing words or wrong grammar.
It is unlikely that these low-quality test cases can contribute to improving NLP software in practice.

Hence, an automatic quality evaluation metric that can help filter out low-quality test cases generated by the existing testing techniques is highly in demand.
Nevertheless, designing an automatic quality evaluation metric for NLP test cases is highly challenging.
First, existing testing criteria are mainly based on coverage metrics, such as code coverage for traditional software \cite{codecoverage} and neuron coverage for deep neural networks \cite{deepxplore}, which cannot be directly leveraged to detect false alarms and evaluate the quality of a natural language test case.
Second, general semantic similarity evaluation metrics fail to detect \textit{Inconsistent} issues under this scenario.
Specifically,
(1) most of the words in the original text and the generated text are the same while existing metrics evaluate the semantic similarity based on all the words in the text and thus, the impact of the mutated word(s) easily vanish;
(2) a word may have different meanings in different contexts, making it difficult to compare only the mutated word(s).
Third, existing work on naturalness evaluation metric either relies on human evaluation \cite{reevaluate2} or qualitative analysis (\textit{e.g.}, part-of-speech checking \cite{bae}), while we need an automatic and quantitative naturalness evaluation metric.

\begin{table*}
\caption{Details of the selected testing techniques.}
\label{tab-techniques}
\begin{center}
\begin{tabular}{c|lll}
\toprule
\bf Technique & \bf Selection & \bf Substitution & \bf Constraints \\
\cmidrule(lr){1-4}
Generative Algorithm (GA) \cite{fast-ga} & k-nearest neighbors & Random combination & \begin{tabular}{@{}l@{}} Percentage of modified words; \\ Euclidean distance; \\ LM grammar checking \\ \end{tabular} \\
\cmidrule(lr){1-4}
BERT-base Adversarial Examples (BAE) \cite{bae} & Word importance & PLM mask prediction & \begin{tabular}{@{}l@{}} Euclidean distance; \\ Part-of-speech checking \\ \end{tabular} \\
\cmidrule(lr){1-4}
Particle Swarm Optimization (PSO) \cite{pso} & Optimization \cite{pso-optimization} & Knowledge graph \cite{hownet} & None \\
\cmidrule(lr){1-4}
Checklist \cite{checklist} & Random selection & \begin{tabular}{@{}l@{}} Transformations: \\ Contraction; Extension; \\ Changing entities \end{tabular} & Numbers of transformation \\
\bottomrule
\end{tabular}
\end{center}
\end{table*}

To address these problems, we introduce {\methodname}, a method for \textbf{A}utomatic \textbf{E}valuation \textbf{O}f \textbf{N}LP test cases.
{\methodname} takes a text pair <original text, generated text> as input and outputs scores regarding semantic similarity and syntactic correctness, aiming for detecting \textit{Inconsistent} and \textit{Unnatural} issues, respectively.
We use {\methodname} to analyze the quality of NLP test cases generated from four popular testing techniques \cite{fast-ga, pso, bae, checklist} on five datasets \cite{mr, agnews-yelp, mnli, snli, glue} which cover three typical NLP tasks, namely natural language inference, sentiment analysis, and semantic equivalence.
We conduct a comprehensive human evaluation on the semantic similarity and language naturalness between the original texts and the generated test cases, and we check whether {\methodname}'s score aligns with human evaluation or not.
The results show that {\methodname} achieves the Average Precision (AP), Area Under Curve (AUC), and Pearson Correlation Coefficient (PCC) scores of 0.688, 0.742, and 0.922, outperforming the best baseline metric by 10\%, 8.1\%, and 7.8\% respectively.
On the evaluation of human judgment of language naturalness, {\methodname} also surpasses all baselines and achieves the average AP, AUC, PCC scores of 0.69, 0.63, 0.82.
These results demonstrate the effectiveness {\methodname} on detecting false alarms and evaluating the language naturalness of NLP test cases.
We also show that the high-quality test cases selected by {\methodname} can significantly improve the accuracy and robustness of NLP software via model training.
Our contributions can be summarized as:
\begin{itemize}
    \item We conduct a comprehensive user study on the test cases generated by existing NLP software testing techniques and find that 85\% of them suffer from two issues: \textit{Inconsistent} and \textit{Unnatural}, resulting in a false alarm rate of 44\%.
    \item We introduce {\methodname}, the first approach to quantitatively evaluate the quality of NLP test cases from semantics and language naturalness, addressing two main quality issues of NLP test cases mentioned above.
    \item {\methodname} is employed to evaluate the test cases generated by four testing techniques on five widely-used datasets, which shows that {\methodname} achieves the best performance in terms of average AP, AUC, and PCC on all datasets.
    \item The implementation of {\methodname}, the raw experimental results, and the human annotation on the test case quality are available on Github\footnote{\url{https://github.com/CUHK-ARISE/AEON}}.
\end{itemize}

%% file: Sections/2_Preliminaries.tex
\section{Preliminaries}
\label{sec-preliminary}

\subsection{Testing Techniques for NLP Software}
\label{sec-testing-techniques}

Though many papers have proposed testing techniques for Computer Vision (CV) software (\textit{e.g.}, face recognition system) \cite{deeptest, deephunter, deepxplore, cw}, the characteristics of natural language make NLP software testing distinguished from that in CV software.
The most significant difference between NLP test cases and CV test cases is that the input space of textual data is not as continuous as images, making every mutation in the original text perceptible.
In addition, in natural language, mutating a single word can cause considerable semantic differences, which further leads to the risk of changing the correct label of the text.
Therefore, when NLP testing techniques assign the label of the original text to the generated test cases, lots of false alarms occur.

Current testing techniques\footnote{In this paper, we consider papers on attacking NLP models as a line of research on testing NLP software because the adversarial examples generated by these techniques can be regarded as test cases for NLP software.} for NLP software can be roughly divided into four categories: character-level, word-level, sentence-level, and multi-level \cite{nlpadv-survey1}.
Character-level techniques \cite{textbugger} mutate a few characters that do not affect human reading comprehension.
Word-level techniques \cite{pwws, checklist} are based on word substitution, usually using synonyms sets or Pre-trained Language Models (PLMs).
Sentence-level techniques \cite{pinjia3} change the whole structure of the sentences either by adding a sentence to the original texts or transforming the entire texts into another semantically similar format.
Those combining different levels of techniques \cite{liang} can be categorized into multi-level techniques.
In particular, word-level techniques significantly outperform others in terms of efficiency \cite{pso}, applicability, and usefulness in robust training \cite{pso, pwws}.
However, this kind of technique suffers more from low-quality test cases \cite{reevaluate2}.
Thus, we focus on test cases generated by word-level testing techniques.

From the perspective of combinatorial optimization, generating test cases with word-level techniques can be formulated as a searching problem, where we substitute each word in the original text to other words in our vocabulary.
The whole search space is the number of words in original text $N$ (where we substitute) times the vocabulary size $V$ (word candidates).
In general, these techniques include diverse modules to prune the search space, which can be classified into three components: target word selection, word substitution, and generation constraints \cite{nlpadv-survey1, textattack}.
Table~\ref{tab-techniques} presents the modules of the four selected testing techniques in terms of the three components.
A suitable target word selection method can decrease $N$ while a proper word substitution method can cut back $V$.
Constraints are commonly applied to ensure that the synthesized texts preserve semantic meaning and are syntactically correct.

\subsection{Problem Definition}
\label{sec-problem-definition}

Given NLP software $\mathcal{F}: \mathcal{X} \rightarrow \mathcal{Y}$ which takes a text $x$ in text space $\mathcal{X}$ as input and outputs its prediction $y \in \mathcal{Y}$, an word-level test case $\hat{x}$ is synthesized from a seed datum $x$ whose ground truth (\textit{i.e.}, label) is $y$ (denoting with $gt(x) = y$).
It also needs to satisfy that $Sim(\hat{x},\ x) \ge c$, where $Sim: \mathcal{X} \times \mathcal{X} \rightarrow \mathbb{R}$ is a similarity metric.
Intuitively, it means $\hat{x}$ and $x$ have equivalent or similar semantics.
$c$ is a task-specific constant to trade between semantic similarity and generation diversity.
Most NLP testing techniques assume that the generated test case and the original text have the same label, \textit{i.e.}, $gt(\hat{x}) = gt(x) = y$.

Given a set of generated test cases $\hat{X}$, sometimes the similarity metric it uses may not be able to detect some inconsistency, then \textit{Inconsistent} test cases occur.
A test case $\hat{x} \in \hat{X}$ is \textit{Inconsistent} when $Sim^{\prime}(\hat{x}, x) \le t_s$.
Here $t_s$ is a threshold, and $Sim^{\prime}$ is a trustable and robust similarity metric, for example, human judgment.
There is a high chance that \textit{Inconsistent} test cases are false alarms, which satisfy $gt(\hat{x}) \neq gt(x)$ and $\mathcal{F}(\hat{x}) = gt(\hat{x})$.
In other words, the software behaves correctly (produces label $gt(\hat{x})$), but we wrongly assume it needs to produce label $gt(x)$.
Whether a test case is \textit{Unnatural} or not depends on human evaluation.
It is defined as $Nat^{\prime}(\hat{x}) \le t_n$ where $Nat^{\prime}: \mathcal{X} \rightarrow \mathbb{R}$ is human judgment on the language naturalness of the textual test case and $t_n$ is a threshold.

\begin{algorithm}[t]
\caption{Algorithm for SemEval}
\label{alg-semeval}
\begin{algorithmic}[1]
\Require Original text $x$; Generated test case $\hat{x}$
\Ensure Semantic similarity score $Sim(x,\ \hat{x}) \in [0,\ 1]$
\State $x \gets$ Tokenize($x$)
\State $\hat{x} \gets$ Tokenize($\hat{x}$)
\State $diff\_indices \gets$ LevenshteinDistance($x,\ \hat{x}$)
\State $emb(x) \gets$ GetWordEmbedding($x$)
\State $emb(\hat{x}) \gets$ GetWordEmbedding($\hat{x}$)
\State $patch\_sim \gets$ a list
\For{each $i \in diff\_indices$}
    \State $patch\_x \gets emb(x)[i - 2:i + 2]$
    \State $patch\_\hat{x} \gets emb(\hat{x})[i - 2:i + 2]$
    \State Append $\frac{patch\_x^T patch\_\hat{x}}{\lVert patch\_x \rVert \cdot \lVert patch\_\hat{x} \rVert}$ to $patch\_sim$
\EndFor
\State $text\_sim \gets \frac{emb(x)^T emb(\hat{x})}{\lVert emb(x) \rVert \cdot \lVert emb(\hat{x}) \rVert}$
\State $min\_sim \gets$ min($patch\_sim$)
\State $avg\_sim \gets$ average($patch\_sim$)
\State $Sim(x,\ \hat{x}) \gets \lambda_1 min\_sim + \lambda_2 avg\_sim + (1 - \lambda_1 - \lambda_2) text\_sim$
\end{algorithmic}
\end{algorithm}

The task of this paper is to design an automatic evaluation metric that can reflect test case quality in terms of semantic consistency and naturalness, which facilitates the detection of \textit{Inconsistent} (false alarms) and \textit{Unnatural} test cases.

%% file: Sections/3_Approaches.tex
\section{Approaches and Implementation}
\label{sec-approaches}

This section introduces the details of {\methodname} whose input is a text pair <original text, generated text> and outputs are a semantic score and a syntactic score.
{\methodname} consists of two parts: \texttt{SemEval} (Semantic Evaluator), which captures the semantic difference between input text pair, and \texttt{SynEval} (Syntactic Evaluator), which assesses how likely the generated test case will be used (\textit{i.e.}, written or typed) by real users.
These two components aim to address \textit{Inconsistent} and \textit{Unnatural} issues, respectively.
In the rest of this section, we will introduce the details of the key components of the two evaluators.

\subsection{SemEval}
\label{sec-semeval}

\texttt{SemEval} aims to solve the two challenges mentioned above.
(1) The influence of the mutated position can easily vanish when taking average since most words in the original text and the generated test case are the same.
(2) Metrics comparing words without contexts can neglect their alternative meanings (\textit{i.e.}, polysemy).
To this end, we propose to combine Levenshtein distance \cite{edit-distance} and sentence embedding model to evaluate the semantic similarity in the NLP testing scenario.
The approach is surprisingly effective considering its simplicity, which is shown in Alg.~\ref{alg-semeval}.

After tokenizing the input texts (line 1-2), which converts all words and punctuation as individual tokens, \texttt{SemEval} extracts small patches of text where the two inputs differ using Levenshtein distance (line 3).
With the help of Levenshtein distance, we can find all mutated positions in linear time.
Next, it applies a PLM to obtain the embeddings of all tokens in the two inputs (line 4-5).
Current PLMs \cite{bert, roberta, infersent, simcse} can be leveraged to project tokens to the embedding space, so this module can be replaced easily when more powerful PLMs are proposed.
Then, for all patches and the whole text, we compute the cosine similarity defined by $\frac{a^T b}{\lVert a \rVert \cdot \lVert b \rVert}$ in line 7-9 and line 12, respectively.
Note that we extract totally five tokens as the patch for a mutation happened in position $i$ by $[i-2:i+2]$.
For a mutation at the beginning or the end of a sentence, we extract the first or last three tokens as our patch.
Finally, we compute the minimum and the average numbers among all patch similarities (line 13-14) and combine them with the text similarity using two hyper-parameters, $\lambda_1$ and $\lambda_2$ (line 15).
After this convex combination, we obtain the output of \texttt{SemEval}, namely $Sim(x,\ \hat{x})$.

\begin{algorithm}[t]
\caption{Algorithm for SynEval}
\label{alg-syneval}
\begin{algorithmic}[1]
\Require Generated test case $\hat{x}$
\Ensure Language naturalness score $Nat(\hat{x}) \in (0,\ 1]$
\State $\hat{x} \gets$ Tokenize($\hat{x}$)
\State $perplexity \gets$ a list
\For{each $token \in \hat{x}$}
    \State $masked\_text \gets$ replace $token$ with \texttt{[MASK]} in $\hat{x}$
    \State $prob \gets$ GetMaskPrediction($masked\_text$)
    \State $prob\_token \gets prob[token]$
    \State Append $prob\_token$ to $perplexity$
\EndFor
\State $min\_nat \gets$ min($perplexity$)
\State $avg\_nat \gets$ average($perplexity$)
\State $Nat(\hat{x}) \gets \phi min\_nat + (1 - \phi) avg\_nat$
\end{algorithmic}
\end{algorithm}

We tackle challenge (1) by considering the minimum and average patch similarities.
For challenge (2), {\methodname} extract the mutated position along with its context, which can improve its ability to understand semantics.
Consider an example:
\begin{table}[H]
    \centering
    \begin{tabular}{lp{5.5cm}}
    \toprule
    \rowcolor{mygray}
    \multicolumn{2}{l}{\textbf{Case Study 1}} \\
    \textsc{Task/Dataset} & SE/QQP \\
    \textsc{Technique} & BAE \\
    \textsc{Original Text} & Is it OK to leave an iPhone plugged into the charger after 100\% \textcolor{red}{charged}? \\
    \textsc{Generated Text} & Is it OK to leave an iPhone plugged into the charger after 100\% \textcolor{red}{indicted}? \\
    \bottomrule
    \end{tabular}
\end{table}

\noindent If we only consider the mutated position \texttt{charged} and \texttt{indicted}, the similarity is high since they are synonyms in the meaning of ``being accused''.
However, \texttt{charged} here means "to put electricity into an electrical device".
This kind of relationship can be captured by its context, which is modeled in the PLMs.

\subsection{SynEval}
\label{sec-syneval}

\begin{table*}
\caption{Details of the selected datasets. NLP tasks include Sentiment Analysis (SA), Natural Language Inference (NLI), and Semantic Equivalence (SE)}
\label{tab-datasets}
\begin{center}
\begin{tabular}{c|ccl}
\toprule
\bf Dataset & \bf Task & \bf Classes & \bf Description \\
\cmidrule(lr){1-4}
Rotten Tomatoes Movie Review (MR) \cite{mr} & SA & 2 & Short sentences or phrases of movie reviews \\
\cmidrule(lr){1-4}
Yelp Restaurant Review (Yelp) \cite{agnews-yelp} & SA & 2 & Long sentences or paragraphs of restaurant reviews \\
\cmidrule(lr){1-4}
Stanford Natural Language Inference (SNLI) \cite{snli} & NLI & 3 & Short texts with simple contexts \\
\cmidrule(lr){1-4}
Multi-Natural Language Inference (MNLI) \cite{mnli} & NLI & 3 & Multi-genre, multi-length texts with complicated contexts \\
\cmidrule(lr){1-4}
Quora Question Pairs (QQP) \cite{glue} & SE & 2 & Two similar questions from Quora \\
\bottomrule
\end{tabular}
\end{center}
\end{table*}

Since synthesized test cases may include grammar errors, punctuation errors, or produce rarely used words and phrases, it is vital to use an automatic and quantitative metric to filter out these \textit{Unnatural} test cases.
Note that this kind of sentence rarely appears in real-world natural languages, hence they are treated as noises and ignored during the training process of PLMs \cite{bert, roberta, albert}.
Intuitively, how natural a sentence is can be reflected by the probability that the sentence has the same distribution as its training data, which can be estimated by PLMs.
Therefore, \texttt{SynEval} is designed to measure naturalness through the perplexity of PLMs.
Perplexity, in its formal definition, is the exponential form of the cross entropy of the given sentence \cite{ppl}, having the form of:
\begin{equation}
\label{eq1}
    perplexity(x) = \sqrt[N]{\prod_{i = 1}^N \frac{1}{P(x_i | x_{1:i - 1})}},
\end{equation}
where $x_i$ is the $i$-th word in the sentence and $x_{1:i - 1}$ is the first to $i - 1$-th words in the sentence.
$perplexity: \mathcal{X} \rightarrow [1,\ \infty)$ measures how confused the PLM is when it sees $x_i$ given $x_{1:i - 1}$, the greater the more confused (\textit{i.e.}, worse).

The recently proposed BERT-like models, including BERT \cite{bert}, RoBERTa \cite{roberta}, and ALBERT \cite{albert} which trained on billions of sentences, are powerful PLMs for modeling this probability.
However, BERT and its variants are bi-directional, taking not only $x_{1:i - 1}$ but also $x_{i + 1:n}$ as input.
Therefore, we need to replace $x_{1:i - 1}$ with $x_{\backslash i}$ in Eq.~\ref{eq1}, where $x_{\backslash i}$ denotes the input sentence with its $i$-th word being \texttt{[MASK]}.
Since our semantic evaluator outputs similarity scores in $(0,\ 1]$ (the greater, the more similar), we adopt $\sqrt[N]{\prod_{i = 1}^N P(x_i | x_{\backslash i})}$ for \texttt{SynEval}, having the same value range of $(0,\ 1]$ (the greater, the better).

Alg.~\ref{alg-syneval} illustrates the implementation of \texttt{SynEval}.
First we tokenize the input (line 1).
Then, for each token in the input, we replace it with the special token \texttt{[MASK]} (line 4).
Feeding the masked text to the PLM, we can obtain the prediction of the masked position, which is a probability distribution over the entire vocabulary (line 5).
Next, we find out the probability that the PLM thinks the masked position can be filled with the original token and record it as the perplexity of this token (line 6-7).
Finally, we compute the minimum and the average numbers among all perplexities and combine them using a hyper-parameter $\phi$.
The score after this convex combination is the output of \texttt{SynEval}, namely $Nat(\hat{x})$.

%% file: Sections/4_Settings.tex
\section{Experimental Design and Settings}

In this paper, we focus on the following four research questions:

\noindent \textbf{RQ1}: What is the quality of the test cases generated by existing testing techniques (Section \ref{sec-rq1})?

\noindent \textbf{RQ2}: How effective is {\methodname} (Section \ref{sec-rq2})?

\noindent \textbf{RQ3}: How can {\methodname} help in testing NLP software? (Section \ref{sec-rq3})

\noindent \textbf{RQ4}: How can {\methodname} help in improving NLP model? (Section \ref{sec-rq4})

\subsection{Testing NLP Software}

To answer the RQs, the first step is generating test cases, \textit{i.e.}, testing NLP software.
We choose to test the APIs provided by Hugging Face Inc.\footnote{\url{https://huggingface.co/}}, the largest NLP open-source community, on five widely-used datasets across three typical tasks: sentiment analysis, natural language inference, and semantic equivalence.

\textbf{Datasets.}
Sentiment analysis aims at classifying the polarity (either positive or negative) of the sentiment of given texts.
The inputs of natural language inference tasks are two pieces of texts, namely Premise and Hypothesis, and the target is to predict whether the Hypothesis is a contradiction, entailment, or neutral to the given Premise.
If the Premise can infer the Hypothesis, the output is entailment; if the Premise can infer NOT Hypothesis, the output is contradiction; otherwise, the output is neutral.
The inputs of semantic equivalence tasks are two pieces of text, namely question 1 and question 2, and the objective is to judge if the meaning of the two given questions is equivalent.
We select five datasets, namely MR, Yelp, SNLI, MNLI, and QQP, for our experiments, whose details are shown in Table~\ref{tab-datasets}.
MR and Yelp are crawled from the internet, so the data contain noises such as HTML tags, HTML encodings, HTML entity names, and hyperlinks, which will make the generated test case hard to read.
To eliminate the influence of noisy data in our human evaluation, we convert HTML texts to plain texts and remove hyperlinks using regular expressions.

\textbf{Testing.}
To be more specific, we choose five BERT-based APIs\footnote{\url{https://huggingface.co/textattack/bert-base-uncased-rotten-tomatoes} \\ \url{https://huggingface.co/textattack/bert-base-uncased-yelp-polarity} \\ \url{https://huggingface.co/textattack/bert-base-uncased-MNLI} \\ \url{https://huggingface.co/textattack/bert-base-uncased-snli} \\ \url{https://huggingface.co/textattack/bert-base-uncased-QQP}} for five different datasets.
According to the statistics given by Hugging Face Inc, these APIs are downloaded more than 30k times every month on average.
Using the testing techniques described in Table~\ref{tab-techniques} implemented by TextAttack \cite{textattack} with their default settings, we generate test cases for all datasets (APIs).
We select 400 original texts for each dataset using each technique, resulting in 8,000 test cases.
After testing the APIs with our test cases, 3,262 test cases (40.8\%) are reported as software errors.

\subsection{Human Evaluation}
\label{sec-human_eval}

We aim to find out whether the reported cases really trigger the erroneous behaviors of NLP software, in other words, whether they are false alarms.
To this end, we design and launch a user study.

\begin{table}
\caption{Details of the selected baselines.}
\label{tab-baselines}
\begin{center}
\begin{tabular}{lcl}
\toprule
& \bf Baseline & \bf Description \\
\cmidrule(lr){1-3}
\multirow{4}{*}{\rotatebox[origin=c]{90}{NC-based \ \ \ }} & NC \cite{deepxplore} & The ratio of activated neurons \\
\cmidrule(lr){2-3}
& NBC \cite{deepgauge} & Activation outside upper/lower bounds \\
\cmidrule(lr){2-3}
& TKNC \cite{deepgauge} & The ratio of top-$k$ activated neurons \\
\cmidrule(lr){2-3}
& BKNC \cite{deephunter} & The ratio of bottom-$k$ activated neurons \\
\cmidrule(lr){1-3}
\multirow{6}{*}{\rotatebox[origin=c]{90}{NLP-based \ \ \ \ \ \ \ }} & BLEU \cite{bleu} & The overlaps of $n$-grams \\
\cmidrule(lr){2-3}
& Meteor \cite{meteor} & $n$-grams with synonyms in WordNet \cite{wordnet} \\
\cmidrule(lr){2-3}
& InferSent \cite{infersent} & BiLSTM-based embedding model \\
\cmidrule(lr){2-3}
& SBERT \cite{sentence-bert} & BERT-based embedding model \\
\cmidrule(lr){2-3}
& SimCSE \cite{simcse} & Embedding model with contrastive learning \\
\cmidrule(lr){2-3}
& BERTScore \cite{bert-score} & Token matching in BERT embedding space \\
\bottomrule
\end{tabular}
\end{center}
\end{table}

\textbf{Design.}
Following \cite{evaluation-survey}, we propose a unified framework to measure the quality of generated test cases.
The quality is defined from four perspectives, including Naturalness, Consistency, Human Label, and Difficulty:
\begin{itemize}[leftmargin=*]

\item Consistency: From ``1 strongly disagree'' to ``5 strongly agree'', how much do you think the two sentences have the same meaning? Consistency quantifies the semantic similarity between the original text and the changed text.

\item Naturalness: From ``1 very bad'' to ``5 very good'', how fluent and natural do you think this sentence is?
Naturalness measures the fluency and grammar of the examples, including grammar errors, punctuation errors, and spelling errors (unrecognizable words).

\item Human label: Ask humans to do the tasks of the given datasets.
It is a task-specific question and records the human judgment of classification answers.

\item Difficulty: From ``1 very easy'' to ``5 very hard'', how difficult for you to make the decision?
Difficulty reflects how difficult the task is for humans.

\end{itemize}

Based on our definition in Sec.~\ref{sec-introduction}, high-quality text cases should have high naturalness and consistency scores.
Human label and difficulty are used to classify the human evaluation results.
We also ask annotators whether these test cases have other problems/issues that we have not identified.
The responses show that \textit{Inconsistent} issue and \textit{Unnatural} issue can cover all their concerns.

\textbf{Crowdsourcing.}
We distribute our questionnaire on Qualtrics\footnote{\url{https://www.qualtrics.com/}}, a platform to design, share, and collect questionnaires.
We recruit crowd workers on Prolific\footnote{\url{https://prolific.co/}}, a platform to post tasks and hire workers.
Since our questions require a high level of reading comprehension and inference skills in English, we require Prolific workers to have a bachelor's degree or above and have English as their first and most fluent language.
Since we focus on false alarms, we randomly sample 100 test cases per dataset that are reported as software errors for human evaluation.
In total, we choose 500 test cases and generate 2,000 questions.
For each question, we ask three workers to give their judgment to reduce the variance.
Therefore, we ask 150 workers to complete all questionnaires.
It takes each worker 15-25 minutes to answer around 40 questions in a questionnaire, and each worker is paid about 5 pounds per hour.
The total cost is 300 pounds.

\subsection{Baselines}
\label{sec-baselines}

We select diverse test case evaluation metrics as baselines from two categories: Neuron Coverage (NC) metrics and NLP-based metrics, which are summarized in Table~\ref{tab-baselines}.

\textbf{NC-based.}
NC and its variants are commonly-used for evaluating test cases.
Different from {\methodname}, NC-based metrics mainly aim at the evaluation of a test set instead of a test case.
In our experiments, we consider NC-based metrics in two ways.
(1) For basic Neuron Coverage (NC) \cite{deepxplore} and Neuron Boundary Coverage (NBC) \cite{deepgauge}, we calculate the NC scores of each generated test case.
(2) For Top-$k$ Neuron Coverage (TKNC) \cite{deepgauge} and Bottom-$k$ Neuron Coverage (BKNC) \cite{deephunter}, they cannot be adapted to a single test case (\textit{e.g.}, TKNC produces the same coverage for single test cases), thus we compute the number of neurons covered by the generated test case but not by the original text.
Intuitively, changes in texts may be reflected in neuron activation.
Note that the comparison with NC-based metrics is not apples-to-apples because NC-based metrics mainly evaluate the quality of a test set.
We include the comparison here for the completeness of our discussion.

\textbf{NLP-based.}
Since the main reason behind false alarms is that the generated test cases cannot keep equivalent or similar semantic meaning with the original text, we include multiple semantic similarity metrics for the baselines of \texttt{SemEval}.
Evaluating the semantic similarity of texts has long been a complex problem in NLP research.
Previous metrics can be divided into corpus-based, knowledge-based, and DNN-based.
DNN-based metrics outperform other methods and have served as a breakthrough in semantic similarity research \cite{semantic-survey}.
We consider a corpus-based metric, BLEU, a knowledge-based metric, Meteor, and four DNN-based metrics, InferSent, SBERT, SimCSE, and BERTScore.
For embedding models, namely InferSent, SBERT, and SimCSE, we report the semantic similarity based on cosine similarity because it is used by most of the researchers \cite{bae, simcse, bert-score} and Euclidean distance yields similar results in all our experiments.





\subsection{Evaluation Criteria}

We compute three criteria: AP, AUC, and PCC, to discover the correlation between human judgment (Sec.~\ref{sec-human_eval}) and the automatic evaluation metrics, including {\methodname}.
We treat the scoring systems as binary classification systems, the human judgment as ground truth, and draw their Precision-Recall curve (P-R curve) and Receiver Operating Characteristic curve (ROC curve) to calculate AP and AUC.
P-R curve shows the trade-off between recall (\textit{i.e.}, true positive rate) and precision, while the ROC curve depicts the trade-off between true positive rate and false positive rate.
AP and AUC represent the area under P-R curve and ROC curve, respectively.
An excellent binary classification system tends to have high AP and AUC scores.
Then, we check whether our scores are correlated with human judgment using PCC, the covariance of two variables divided by the product of their standard deviations, which can be written in the form of:
\begin{equation}
    PCC(X,\ Y) = \frac{Cov(X,\ Y)}{\sqrt{Var(X) \cdot Var(Y)}}.
\end{equation}
PCC is able to show how linearly correlated two variables are.
Note that negative PCC value indicates that the two variables are negatively correlated.

%% file: Sections/5_Results.tex
\section{Experimental Results}

\begin{figure}
  \centerline{\includegraphics[width=0.82\linewidth]{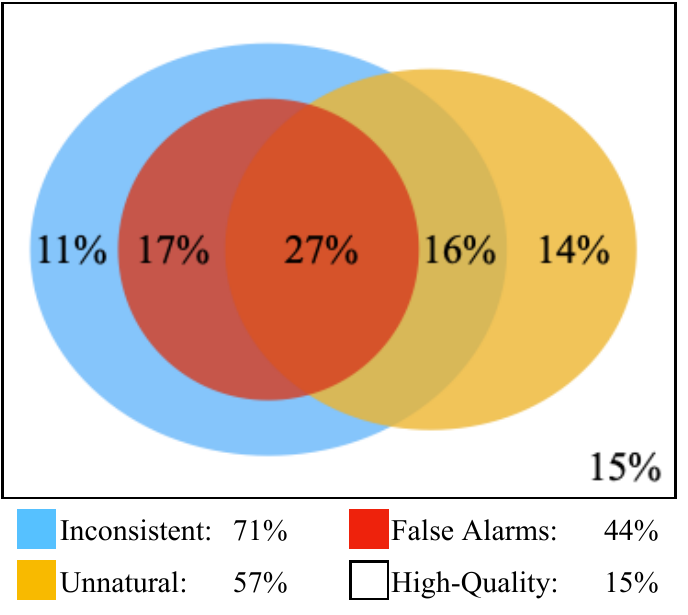}}
  \caption{
  Venn Diagram of the proportion of each vulnerability category (better viewed in colored mode).
  }
  \label{fig-venn}
\end{figure}

\subsection{RQ1: The Quality of Test Cases}
\label{sec-rq1}

We average the consistency, naturalness, and difficulty scores as the respective final scores.
We use the label that most workers agree as the final human label.
For each test case, we decide whether it is an \textit{Inconsistent} case or an \textit{Unnatural} case based on the consistency and naturalness scores.
Finally, if the human label differs from the given label, the test case is considered a false alarm.

\begin{table*}
\caption{AP, AUC, and PCC results that show how well the automatic metrics align with human-annotated consistency scores.}
\begin{center}
\begin{tabular}{ll ccc ccc ccc ccc ccc}
\toprule
& \bf Datasets &
\multicolumn{3}{c}{\bf MR} &
\multicolumn{3}{c}{\bf Yelp} &
\multicolumn{3}{c}{\bf SNLI} &
\multicolumn{3}{c}{\bf MNLI} &
\multicolumn{3}{c}{\bf QQP} \\
                   \cmidrule(lr){3-5} \cmidrule(lr){6-8} \cmidrule(lr){9-11} \cmidrule(lr){12-14} \cmidrule(lr){15-17}
& \bf Metrics      & AP & AUC & PCC & AP & AUC & PCC & AP & AUC & PCC & AP & AUC & PCC & AP & AUC & PCC \\
\midrule
\multirow{4}{*}{\rotatebox[origin=c]{90}{NC-based \ \ \ }} &
\bf NC             & 0.76 & 0.55 & 0.73 & 0.29 & 0.36 & -0.70 & 0.56 & 0.55 & 0.65 & 0.31 & 0.54 & 0.50 & 0.44 & 0.52 & 0.47 \\
\cmidrule(lr){2-17}
& \bf NBC          & 0.78 & 0.66 & 0.85 & 0.51 & 0.62 & 0.85 & 0.44 & 0.47 & 0.23 & 0.35 & 0.56 & 0.90 & 0.38 & 0.56 & 0.78 \\
\cmidrule(lr){2-17}
& \bf TKNC         & 0.88 & 0.73 & 0.88 & 0.56 & 0.64 & 0.88 & 0.58 & 0.57 & 0.65 & 0.47 & 0.66 & 0.87 & 0.67 & 0.76 & 0.91 \\
\cmidrule(lr){2-17}
& \bf BKNC         & 0.88 & 0.73 & 0.86 & 0.55 & 0.63 & 0.85 & 0.58 & 0.58 & 0.67 & 0.48 & 0.67 & 0.87 & 0.64 & 0.75 & 0.88 \\
\cmidrule(lr){1-17}
\multirow{6}{*}{\rotatebox[origin=c]{90}{NLP-based \ \ \ \ \ \ \ }} &
\bf BLEU           & 0.75 & 0.58 & 0.75 & 0.45 & 0.57 & 0.45 & 0.45 & 0.42 & -0.36 & 0.34 & 0.50 & 0.44 & 0.44 & 0.54 & 0.44 \\
\cmidrule(lr){2-17}
& \bf Meteor       & 0.74 & 0.54 & 0.73 & 0.43 & 0.57 & 0.41 & 0.51 & 0.50 & -0.18 & 0.46 & 0.59 & 0.80 & 0.53 & 0.62 & 0.81 \\
\cmidrule(lr){2-17}
& \bf InferSent    & 0.81 & 0.68 & 0.96 & 0.55 & 0.67 & 0.88 & 0.55 & 0.61 & 0.55 & 0.40 & 0.61 & 0.93 & 0.48 & 0.59 & 0.83 \\
\cmidrule(lr){2-17}
& \bf SentBERT     & 0.85 & 0.71 & 0.92 & 0.53 & 0.63 & 0.83 & 0.65 & 0.65 & 0.66 & 0.46 & 0.63 & 0.74 & 0.52 & 0.59 & 0.76 \\
\cmidrule(lr){2-17}
& \bf SimCSE       & 0.85 & 0.75 & 0.93 & 0.63 & 0.76 & 0.92 & 0.61 & 0.65 & 0.76 & 0.54 & 0.73 & 0.96 & 0.50 & 0.56 & 0.68 \\
\cmidrule(lr){2-17}
& \bf BERTScore    & 0.75 & 0.61 & 0.91 & 0.44 & 0.53 & 0.37 & 0.59 & 0.60 & 0.20 & 0.34 & 0.58 & 0.73 & 0.46 & 0.59 & 0.67 \\
\midrule
& \bf AEON (Ours)  & \bf 0.92 & \bf 0.82 & \bf 0.93 & \bf 0.75 & \bf 0.84 & \bf 0.91 & \bf 0.66 & \bf 0.70 & \bf 0.85 & \bf 0.58 & \bf 0.75 & \bf 0.98 & \bf 0.56 & \bf 0.63 & \bf 0.90 \\
\bottomrule
\end{tabular}
\label{tab-semantic}
\end{center}
\end{table*}

To show how severe the problem in NLP test case quality is, we draw the Venn diagram of the generated test cases based on the human annotation results.
As is shown in Fig.~\ref{fig-venn}, 44\% of them change the label and thus are false alarms.
In other words, there are only 1,435 cases triggering software errors in all 8,000 test cases.
57\% of them are not natural enough, while 71\% fail to preserve the semantic meaning.
Only 15\% of them have good language naturalness and preserve the semantic meaning, which are counted as high-quality test cases.
Besides the statistical information, we have two more observations.
First, though the majority of \textit{Inconsistent} cases are false alarms, a few \textit{Inconsistent} cases do not change the label.
These test cases only account for 11\%, and the rest can be categorized into the two issues.
Second, bad naturalness can sometimes hurt semantic meaning, resulting in test cases that are both \textit{Inconsistent} and \textit{Unnatural}.
This is because the unnatural part can eliminate some key information in texts and further change the semantics.

\begin{tcolorbox}[width=\linewidth, boxrule=0pt, top=1pt, bottom=1pt, left=1pt, right=1pt, colback=gray!20, colframe=gray!20]
\textbf{Answer to RQ1:}
The quality of NLP test cases cannot be guaranteed by existing testing techniques.
71\% and 57\% of test cases generated by existing NLP testing techniques are \textit{Inconsistent} and \textit{Unnatural}, respectively.
44\% of test cases are false alarms, significantly degrading the effectiveness and efficiency of existing testing techniques.
\end{tcolorbox}

\begin{table*}
\caption{AP, AUC, and PCC results that show how well the automatic metrics align with human-annotated naturalness scores.}
\begin{center}
\begin{tabular}{l ccc ccc ccc ccc ccc}
\toprule
\bf Datasets &
\multicolumn{3}{c}{\bf MR} &
\multicolumn{3}{c}{\bf Yelp} &
\multicolumn{3}{c}{\bf SNLI} &
\multicolumn{3}{c}{\bf MNLI} &
\multicolumn{3}{c}{\bf QQP} \\
                \cmidrule(lr){2-4} \cmidrule(lr){5-7} \cmidrule(lr){8-10} \cmidrule(lr){11-13} \cmidrule(lr){14-16}
\bf Metrics     & AP & AUC & PCC & AP & AUC & PCC & AP & AUC & PCC & AP & AUC & PCC & AP & AUC & PCC \\
\midrule
\bf NC          & 0.66 & 0.56 & 0.53 & 0.64 & 0.41 & -0.70 & 0.45 & 0.54 & -0.09 & 0.56 & 0.51 & 0.27 & 0.56 & 0.52 & 0.03 \\
\cmidrule(lr){1-16}
\bf NBC         & 0.72 & 0.57 & 0.32 & 0.70 & 0.47 & 0.51 & 0.32 & 0.52 & 0.66 & 0.43 & 0.36 & -0.91 & 0.52 & 0.47 & -0.61 \\
\cmidrule(lr){1-16}
\bf TKNC        & 0.68 & 0.56 & 0.57 & 0.67 & 0.53 & 0.49 & 0.36 & 0.51 & 0.09 & 0.55 & 0.47 & -0.33 & 0.65 & 0.57 & 0.51 \\
\cmidrule(lr){1-16}
\bf BKNC        & 0.67 & 0.55 & 0.52 & 0.66 & 0.52 & 0.45 & 0.36 & 0.51 & 0.09 & 0.55 & 0.47 & -0.39 & 0.64 & 0.55 & 0.46 \\
\midrule
\bf AEON (Ours) & \bf 0.87 & \bf 0.59 & \bf 0.66 & \bf 0.77 & \bf 0.58 & \bf 0.84 & \bf 0.52 & \bf 0.63 & \bf 0.75 & \bf 0.54 & \bf 0.65 & \bf 0.98 & \bf 0.73 & \bf 0.68 & \bf 0.87 \\
\bottomrule
\end{tabular}
\label{tab-syntactic}
\end{center}
\end{table*}

\subsection{RQ2: The Effectiveness of {\methodname}}
\label{sec-rq2}

Since {\methodname} is designed to evaluate the semantic similarity and language naturalness of NLP software test cases, we assess the two modules, \texttt{SemEval} and \texttt{SynEval}, to validate the effectiveness of our approach.
We use default settings for all baselines, and we select $k=192$ (one-fourth of neurons in each layer) for TKNC and BKNC.
We set $\lambda_1 = 0.1,\ \lambda_2 = 0.2$ for \texttt{SemEval}, and $\phi = 0.6$ for \texttt{SynEval}.

\subsubsection{SemEval}

We draw P-R curves and ROC curves for the semantic scores calculated by \texttt{SemEval} as well as the other baselines mentioned in Sec.~\ref{sec-baselines} and consistency from human evaluation.
Then we compute AP and AUC scores, which are shown in Table~\ref{tab-semantic}.
Our method achieves higher AP and AUC values averaged on all datasets and baselines, showing the strong ability of \texttt{SemEval} to filter out \textit{Inconsistent} cases.
The results also validate the effectiveness of \texttt{SemEval} on capturing subtle semantic changes.
We calculate PCC between the semantic score and human-annotated consistency for each method.
As shown in Table~\ref{tab-semantic}, our approach achieves about $0.92$ PCC on average, which significantly outperforms all the baselines.
This shows that the score of the \texttt{SemEval} aligns well with human evaluation.

NC-based metrics achieve decent performance and surpass many NLP-based metrics, especially on MNLI and QQP datasets, indicating that neuron activation patterns can reflect text semantic changes.
As for NLP-based metrics, BLEU and Meteor perform the worst since they cannot handle highly overlapped texts.
The BLEU and Meteor scores for text pairs are always high since most of the words in the original texts and the generated test cases are the same.
DNN-based metrics cannot perform well because of three main reasons.
(1) Word embeddings usually lack semantic information.
For instance, the embeddings of \texttt{[reject]} and \texttt{[accept]} calculated by BERT \cite{bert} have high cosine similarity of $0.846$, while such word substitution changes the correct label in sentiment analysis.
Another example is that the cosine similarity between embeddings of \texttt{[Tom]} and \texttt{[Jack]} is $0.978$, which hurts in natural language inference tasks.
(2) Baselines that employ token matching \cite{bert-score} are prone to mistakenly matching multiple words to a single word.
Consider the following example:
\begin{table}[H]
    \centering
    \begin{tabular}{lp{5.5cm}}
    \toprule
    \rowcolor{mygray}
    \multicolumn{2}{l}{\textbf{Illustrative Example 1}} \\
    \textsc{Original Text} & I do like the movie, though I did not watch it at cinema. \\
    \textsc{Generated Text} & I do \textcolor{red}{not} like the movie, though I did not watch it at cinema. \\
    \bottomrule
    \end{tabular}
\end{table}

\noindent The word \texttt{[not]} in the generated text can be matched to the second \texttt{[not]} in the original sentence, resulting in a high similarity score.
(3) Models based on contrastive learning fail due to the lack of data with subtle differences in their training set.
These models are mainly trained on natural language inference datasets \cite{infersent, simcse}, which can hardly cover the cases where two sentences have few but vital differences.

Considering different datasets, in sentiment analysis tasks, {\methodname} and all other baselines perform better on MR than on Yelp since the texts in MR in shorter and simpler.
For natural language inference tasks, though MNLI is more complicated than SNLI, it is surprising to observe that SNLI has lower PCC scores than MNLI.
There are negative PCC scores when using BLEU and Meteor, indicating the negative correlation between the baselines and human evaluation.
We think the reason behind this is that the complexity of MNLI lies in the diversity of contexts, and changes in contexts typically will not change the corresponding labels.

\subsubsection{SynEval}

To evaluate the performance of \texttt{SynEval}, we draw P-R curves and ROC curves and compute AP and AUC, treating \texttt{SynEval} as a binary classifier to recognize \textit{Unnatural} cases.
We also calculate PCC between \texttt{SynEval} and naturalness score from human evaluation, which is included in Table~\ref{tab-syntactic}.
We can observe that though NC-based metrics have an excellent performance on detecting \textit{Inconsistent} test cases, they fall short of measuring language naturalness.
The performance varies significantly on different datasets.
The PCC scores of NBC, TKNC, and BKNC show a negative correlation on MNLI, while positively correlated on other datasets.
We infer the reason behind this may be that the models make the decision based on the appearance of certain words or phrases, ignoring whether the input texts have good language naturalness.
In addition to using BERT \cite{bert} in \texttt{SynEval} that is presented in Table~\ref{tab-syntactic}, we also try other language models including RoBERTa \cite{roberta} and ALBERT \cite{albert}, among which BERT achieves the highest AUC and AP values, averaged on all datasets.
Note that traditional grammar checkers are not suitable for this task because they do not provide quantitative results, and they cannot reveal the error-free yet strange sentences that people rarely write.

\textbf{The impact of the hyperparameters.}
If we set the proportion of the minimal semantic score from 0 to 1, we can observe that the performance increases at first, then remains stable at the same level, and finally drop when it gets close to 1.
We balance this trade-off using a grid search for lambdas and phis.
These parameters can be generalized to other datasets and NLP tasks since we adopt the same parameters and consistently achieve good performance for all selected datasets in our experiments.
We also test different patch lengths $l$ for extracting $[i-l:i+l]$.
In particular, $l=1$ does not work well because most NLP models use BPE (Byte Pair Encoding) \cite{bpe} for tokenization, which may divide a word into smaller tokens, making it extract only part of a word.
Long patches (e.g., $l\ge5$) suffer from the same problem as average scores, \textit{i.e.}, the impact of mutation vanishes after averaging.
In our experiments, $l=3$ and $l=4$ lead to similar results.
To reduce computation cost, we select $l=4$ for this parameter.

\begin{tcolorbox}[width=\linewidth, boxrule=0pt, top=1pt, bottom=1pt, left=1pt, right=1pt, colback=gray!20, colframe=gray!20]
\textbf{Answer to RQ2:}
{\methodname}, which consists of \texttt{SemEval} and \texttt{SynEval}, is effective in terms of detecting test cases that change the label (false alarms) and test cases that are unnatural.
{\methodname} outperforms all baselines in average on all datasets.
\end{tcolorbox}

\subsection{RQ3: Test Case Selection Using {\methodname}}
\label{sec-rq3}

This paper aims to propose a metric that facilitates NLP software testing by evaluating the quality of test cases.
In this section, we utilize {\methodname} to filter out low-quality test cases.
We conduct experiments to verify whether the test cases selected by {\methodname} enjoy better semantic consistency and language naturalness.
Specifically, {\methodname} can be utilized to filter out \textit{Inconsistency} and \textit{Unnatural} test cases to improve the quality of test cases in average.
For \texttt{SemEval}, we set different thresholds for different tasks.
We choose multiple thresholds for semantic similarity score because whether the label will change depends on the given task.
Consider this pair of texts (original and generated) which is inconsistent in semantics:
\begin{table}[H]
    \centering
    \begin{tabular}{lp{5.5cm}}
    \toprule
    \rowcolor{mygray}
    \multicolumn{2}{l}{\textbf{Illustrative Example 2}} \\
    \textsc{Original Text} & I watched the movie at \textcolor{red}{home}, it was nice. \\
    \textsc{Generated Text} & I watched the movie at \textcolor{red}{cinema}, it was nice. \\
    \bottomrule
    \end{tabular}
\end{table}

\noindent If they are in a sentiment analysis dataset, the label remains unchanged (\textit{i.e.}, positive).
However, if they appear in a natural language inference dataset as premises, and the hypothesis is ``I went out for the movie'', the label changes from contradiction to entailment.
Therefore, to best filter out those false alarms, we set thresholds as 0.87, 0.90, and 0.91 for sentiment analysis, natural language inference, and semantic equivalence, respectively.
The thresholds are computed with a balance between true positive rate and false positive rate.
From the thresholds, we can see that the three tasks need more semantic similarity increasingly to ensure the preservation of labels, which aligns with the characteristics of the datasets.
For language naturalness, we set the threshold as 0.21.

\begin{table}
\caption{The quality of test cases without and with {\methodname}.}
 \begin{center}
\begin{tabular}{l c c c}
\toprule
\bf Data Source       & \bf Consistency & \bf Naturalness & \bf False Alarms \\
\cmidrule(lr){1-4}
\bf w/o {\methodname} & 2.627 & 2.916 & 0.440 \\
\cmidrule(lr){1-4}
\bf w/ {\methodname}  & 3.357 & 3.305 & 0.262 \\
\midrule
\bf Improvement       & $\uparrow 27.8\%$ & $\uparrow 13.3\%$ & $\downarrow 40.6\%$ \\
\bottomrule
\end{tabular}
\label{tab-quality-improvement}
 \end{center}
\end{table}

We generate 500 test cases that are reported to trigger some software errors, including various datasets and testing techniques mentioned in Table \ref{tab-datasets} and Table \ref{tab-techniques} respectively.
Then we check whether human evaluation has improved before and after applying {\methodname} to select high-quality test cases.
The results are shown in Table~\ref{tab-quality-improvement}.
The average consistency and naturalness scores of the 500 test cases are 2.627 and 2.916, which are below 3 (considered as \textit{Inconsistency} and \textit{Unnatural} test cases) in average.
The false alarm rate is 0.44.
After selecting test cases whose \texttt{SemEval} and \texttt{SynEval} scores are above the thresholds with the help of {\methodname}, the quality of test cases is significantly enhanced.
The scores increase to 3.357 and 3.305, considered high-quality test cases on average.
The consistency score improves by 27.8\%, and the naturalness score improves by 13.3\%.
The false alarm rate is 26.2\%, showing a significant improvement of 40.6\%.
The results demonstrate the effectiveness of {\methodname} on high-quality test case selection.

\subsubsection{Case Study}

We choose one of the generated test cases as an example to illustrate the performance of our \texttt{SemEval} and \texttt{SynEval} compared to other baselines.
\begin{table}[H]
    \centering
    \begin{tabular}{lp{5.5cm}}
    \toprule
    \rowcolor{mygray}
    \multicolumn{2}{l}{\textbf{Case Study 2}} \\
    \textsc{Task/Dataset} & SA/MR \\
    \textsc{Technique} & PSO \\
    \textsc{Original Text} & The result is a \textcolor{red}{powerful}, naturally dramatic piece of low-budget filmmaking. \\
    \textsc{Generated Text} & The result is a \textcolor{red}{terrible}, naturally dramatic piece of low-budget filmmaking. \\
    \bottomrule
    \end{tabular}
\end{table}

\noindent {\methodname} achieves a semantic score of 0.58 and a syntactic score of 0.22.
From the semantic side, the sentiment of the original example is positive.
However, the sentiment of the generated test case is negative because \texttt{[terrible]} is a negative adjective.
This test case is not only \textit{Inconsistent} but also a false alarm since the label of this example is changed.
Therefore, an excellent semantic metric should give this test case a low score to filter it out.
Our method, \texttt{SemEval}, gives the text pair a score of 0.58, which is far below the threshold of 0.87, indicating that the test cases cannot preserve the semantics and should be filtered out.
{\methodname} works effectively because our design to consider patch similarity identifies that the substitution (\texttt{[powerful]}$\rightarrow$\texttt{[terrible]}) dramatically changes the semantic meaning.
From the syntactic side, the generated test case reads smoothly without difficulty comprehending its meaning, suggesting that it has good language naturalness.
The case obtains a score of 0.22 given by \texttt{SynEval}, which is above the threshold of 0.21 and indicates that the test case is not an \textit{Unnatural} case.
All in all, our method outperforms other baselines both in semantic and syntactic perspectives on this example.

\begin{tcolorbox}[width=\linewidth, boxrule=0pt, top=1pt, bottom=1pt, left=1pt, right=1pt, colback=gray!20, colframe=gray!20]
\textbf{Answer to RQ3:}
{\methodname} can effectively filter out low-quality test cases.
The remaining test cases enjoy better semantic consistency and language naturalness and, most importantly, a lower false alarm rate.
Thus, {\methodname} can facilitate NLP software testing by selecting high-quality test cases, saving developers' time.
\end{tcolorbox}

\subsection{RQ4: Improving NLP Software with {\methodname}}
\label{sec-rq4}

Although NLP software testing is a promising research direction, it incurs an important yet unavoidable question: can the test cases be utilized to improve NLP software?
To further show how high-quality test cases selected by {\methodname} can help in improving NLP software, we add test cases that the model misclassifies to the training set and conduct model re-training.
Accuracy is verified on the test set of the given task, while robustness is evaluated using the success rate of adversarial attacks.
In this section, we run experiments to verify whether {\methodname} can further improve the robustness and accuracy in model re-training.

We focus on fine-tuning a pre-trained BERT on MR dataset for sentiment analysis task for simplicity and better reproducibility.
We first generate the test cases using the testing techniques mentioned in Table \ref{tab-techniques} using the entire training set of MR for seeds (original texts).
Then we consider two settings:
(1) randomly select as many test cases as 5\% to 25\% cases in the training set to train the model;
(2) rank all the test cases with {\methodname} in descending order and select the same size as (1) to train the model.
After training, which takes five epochs to reach convergence, we evaluate the models' accuracy using the MR test set and robustness using an adversarial attack method, PWWS \cite{pwws}.

\begin{figure}
  \centerline{\includegraphics[width=\linewidth]{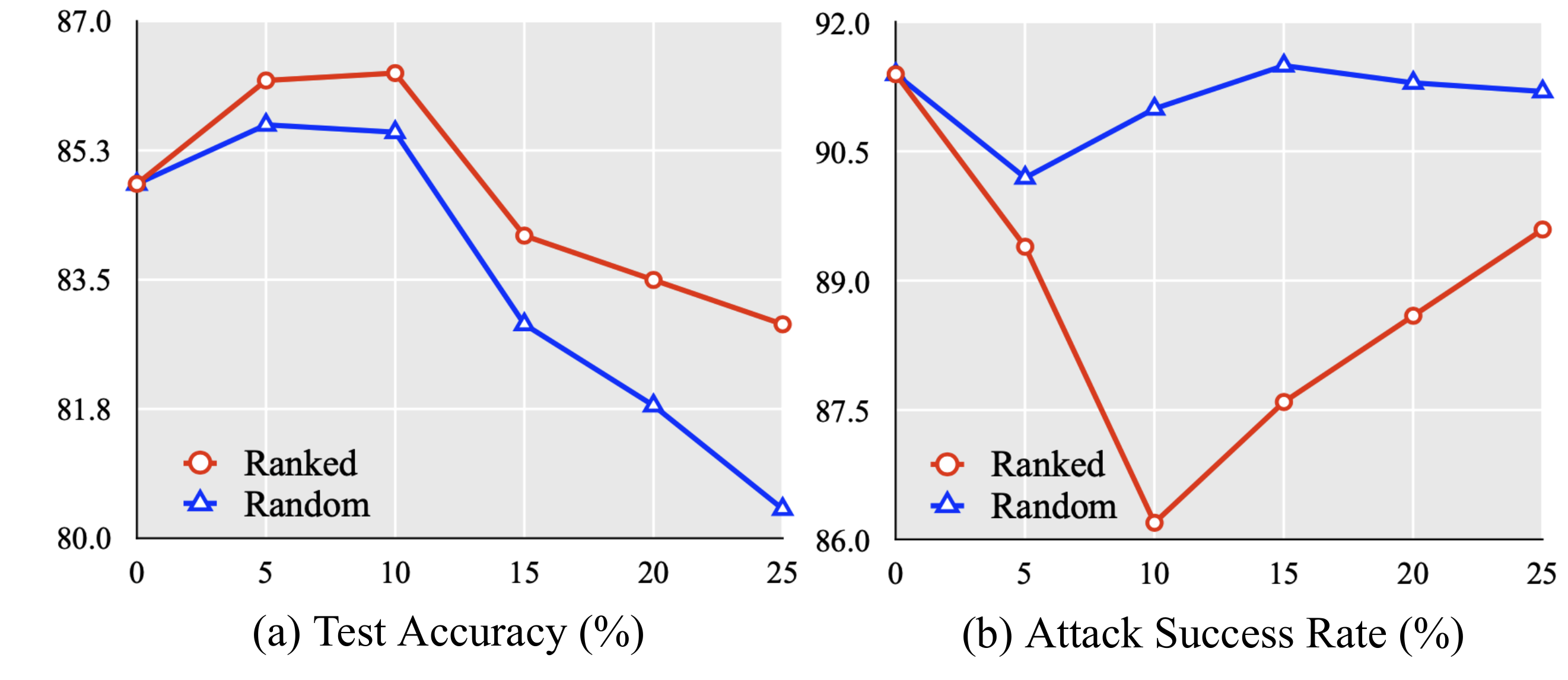}}
  \caption{
  The left figure shows the accuracy of the test set (the higher, the better), and the right figure shows the attack success rate (the lower, the better).
  Blue lines are the results from models trained on randomly selected test cases, while red lines are the results from models trained on test cases selected by {\methodname}.
  The horizontal axis is the ratio of the test cases in the whole training set.
  }
  \label{fig-robust}
\end{figure}

As shown in Fig.~\ref{fig-robust}(a), models trained with ranked test cases outperform the models with randomly selected test cases in terms of accuracy.
In addition, Fig.~\ref{fig-robust}(b) shows that models trained with ranked test cases are more robust (the lower the attack success rate is, the more robust the model is).
At the beginning of each figure, we observe improvement in accuracy and robustness on both lines.
This is because test cases add generalization ability to models.
As we use more additional data in training, the noisy data issue surfaces (\textit{i.e.}, false alarms and low-quality test cases) and starts to harm the model accuracy and robustness.
We can observe that no matter what ratio of test cases we use, the models trained with {\methodname} ranked data achieve higher accuracy and robustness thanks to the high-quality test cases it selects.
In contrast, the accuracy and robustness drop quickly in models trained with randomly selected test cases.
In conclusion, adding low-quality test cases can easily hurt both accuracy and robustness, while adding a reasonable amount of high-quality test cases selected by {\methodname} leads to model improvement in terms of both accuracy and robustness.

\begin{tcolorbox}[width=\linewidth, boxrule=0pt, top=1pt, bottom=1pt, left=1pt, right=1pt, colback=gray!20, colframe=gray!20]
\textbf{Answer to RQ4:}
Equipped with {\methodname}, test cases generated by existing testing techniques can be prioritized.
The prioritized test cases can be further leveraged into NLP model training, resulting in higher accuracy and stronger robustness.
{\methodname} can also save time and resources by filtering out false alarms and test cases of low quality.
\end{tcolorbox}

\subsection{Discussion}

\textbf{The validity of out user study.}
To alleviate workers' negligence in the annotation process, we double-check the results in two cases.
(1) The workers feel it challenging to select one specific label.
Specifically, we check test cases with high difficulty scores (above 3.5), or where annotators returned three different labels (\textit{e.g.}, three workers give the label of contradiction, entailment, and neutral in natural language inference tasks).
(2) The human label is different from the original label (\textit{i.e.}, label changes), while the human-annotated consistency score is still high, which is counter-intuitive since if the label is changed, then the semantic meaning must have changed.
We find out 21 cases for case (1) and 66 examples for case (2) and hire two annotators with two years of experience in NLP research to re-evaluate them.
Afterward, we compute the Kappa score to check inter-rater reliability.
The Fleiss' Kappa of the classification task is 0.76 averaged on all datasets, implying substantial agreement among the annotators.
We do not apply Cohen's Kappa since we have three annotators for each case.

\textbf{The importance of \textit{Unnatural} test cases.}
We evaluate the naturalness of test cases and filter out \textit{Unnatural} ones because they are rarely seen in real-world scenarios.
Though previous work intentionally add spelling and grammar errors to data to improve the robustness of the NLP models \cite{robustencodings}, we noticed annotators mentioned that they could not understand the weird grammar or expression of some generated texts.
In addition, unnaturalness can hide important semantic information, which changes the semantics and further renders the possibility of generating false alarms.
Consider the following example:
\begin{table}[H]
    \centering
    \begin{tabular}{lp{5.5cm}}
    \toprule
    \rowcolor{mygray}
    \multicolumn{2}{l}{\textbf{Case Study 3}} \\
    \textsc{Task/Dataset} & SA/MR \\
    \textsc{Technique} & PSO \\
    \textsc{Original Text} & ... is a relationship that is \textcolor{red}{worthy} of our respect. \\
    \textsc{Generated Text} & ... is a relationship that is \textcolor{red}{costly} of our respect. \\
    \bottomrule
    \end{tabular}
\end{table}
\noindent In this example, the unnaturalness changes the semantics of the sentence and further reverses from positive sentiment to negative sentiment while making readers confused.
Thus, we think \textit{Unnatural} issues are important, and they can also degrade the quality of NLP software test cases.

\textbf{Limitations of our proposed approaches.}
Our proposed approaches consist of two parts, \texttt{SemEval} for semantic semilarity and \texttt{SynEval} for language naturalness.
To apply \texttt{SemEval}, we require both the generated test cases and the original texts at the same time, thus limiting \texttt{SemEval} to testing techniques of metamorphic testing, where the metamorphic relation assumes the transformation will change or will not change the semantics of the original text.
Under the assumption that the transformation will not change the semantics, we filter out those with lower \texttt{SemEval} scores, while under the assumption that the transformation will change the semantics, we filter out those with higher \texttt{SemEval} scores.
\texttt{SynEval} can be applied to any textual test case.

%% file: Sections/6_Related_Work.tex
\section{Related Work}
\label{sec-related-work}

\subsection{Testing AI Software}

With the improvement of Artificial Intelligence (AI) models, companies tend to deploy AI in real-world applications like autonomous driving and neural machine translation \cite{pouyanfar2018survey}.
However, AI software inherits the deficiencies of AI models that they are prone to erroneous behavior given particular inputs \cite{athalye2018obfuscated, carlini2016hidden, cw, du2020sirenattack, fgsm, xiang2019generating, weibin1, weibin2}.
A line of research has been conducted to test AI software systems to address this problem.
Specifically, they test software based on convolutional neural network and feed forward neural network \cite{henriksson2019towards, gambi2019automatically, kim2019guiding, ma2018deepmutation, deeptest, deeproad, weibin3, jianping1}, software based on recurrent neuron network \cite{du2019deepstellar}, and software based on general DNN models \cite{hu2019deepmutation++, zhang2020deepsearch, zhang2020machine}.
Other researchers focus on testing deep learning libraries \cite{pham2019cradle}, assist the debugging process \cite{ma2018mode}, and detect adversarial examples online \cite{ma2019nic, tao2018attacks, wang2019adversarial, xu2017feature}.
Unlike these papers that primarily focus on CV software, this paper focuses on NLP software.

\subsection{Testing NLP Software}

DNNs have boosted the performance of many NLP fields such as code analysis \cite{alon2019code2vec, iyer2016summarizing, pradel2018deepbugs} and machine translation \cite{transformer}.
In recent years, researchers have proposed a variety of metamorphic testing techniques for NLP software \cite{checklist, pinjia1, pinjia2, pinjia3, chen2021validation, chen2021testing, sun2020automatic}.
In addition to metamorphic testing techniques, another line of research for finding NLP software errors \cite{textbugger, textfooler, pso, fast-ga, clare, vlattack} is inspired by the adversarial attack concept in the CV field.
Our work focuses on automatic quality evaluation of test cases generated by these testing techniques.
Thus, we believe {\methodname} complements with existing work.

\subsection{Testing Criteria}
Testing criterion, such as code coverage, has been widely utilized to measure how good a test suite is in traditional software (\textit{e.g.}, compilers).
Inspired by code coverage in traditional software, DeepXplore \cite{deepxplore} introduces the concept of neuron coverage for AI software: the percentage of neurons activated by the test cases.
In recent years, researchers have proposed diverse variants of neuron coverage as testing criteria focusing on different activation magnitudes \cite{deepgauge, deephunter}.
Researchers also develop neuron coverage specially designed for recurrent neuron networks \cite{du2019deepstellar, guo2019rnn-tset, huang2019coverage} to adopt the properties of sequence inputs.
Different from neuron coverage metrics, which often act as test adequacy criteria of the test suite, our approach focuses on the quality evaluation of every test case. 
Thus, we think {\methodname} can complement with existing testing criteria and contribute to research on NLP software testing.

%% file: Sections/7_Conclusion.tex
\section{Conclusion}

This paper is the first to explore the quality of test cases generated by NLP software testing techniques.
In an evaluation study, we surprisingly observe that 44\% of the generated NLP test cases are of low quality, incurring \textit{Inconsistent} or/and \textit{Unnatural} issues.
Thus, instead of improving NLP software, utilization of these test cases in model training could even degrade its accuracy and robustness.
To this end, we introduce {\methodname}, a novel, effective approach for automatic quality evaluation of NLP software testing cases.
Given an original text and a generated test case, {\methodname} returns two scores regarding similarity consistency and language naturalness.
Our evaluation and user study show that {\methodname}'s scores align well with the quality scores returned by humans.
In particular, {\methodname} achieves 69.4\% and 68.5\% AP on detecting \textit{Inconsistent} and \textit{Unnatural} test cases generated by four SOTA testing techniques on five widely-used datasets.
We can conduct test selection or prioritization according to the scores returned by {\methodname}.
In our evaluation, models trained on the test cases selected by {\methodname} consistently achieve better accuracy and robustness than models trained on randomly selected test cases.
We believe that this work is the important first step toward systematic quality evaluation of NLP software test cases, which can further enhance the effectiveness of testing techniques for NLP software and complement existing testing criteria.
We leave the work for automatically fixing \textit{Inconsistent} and \textit{Unnatural} test cases for future exploration.